# Investigation of the electronic properties of the surface and bulk forms of gold and palladium

**U.N. Kurelchuk, P.V. Borisyuk , O.S. Vasilyev, Yu. Yu. Lebedinsky**

National Research Nuclear University MEPhI, Kashirskoye sh., 31, Moskva, 115409

E-mail: unkurelchuk@mephi.ru

**Abstract.** The density of electronic states for bulk metals Au and Pd, their surfaces in the form of polycrystalline surface layers of nanometer thickness is investigated. The calculations were performed using density functional theory with pseudopotential in full relativistic approximation. Approximations have been found that provide calculations the density of electronic states of noble metal surfaces that describe the experimentally observed features of XPS spectra of the valence band of these metals.

**Introduction**

Nanoporous films made of noble metallic nanoclusters are promising materials for high-efficiency thermoelectric elements. Modern computational modeling and numeric methods allow solving such problems as: prediction of stable nanoclusters from, predetermined materials with certain dimensions, theoretical study of their electronic, thermal, magnetic and other properties; computational experiments for planning and optimizing physical experiments and interpreting their results.

In this article we consider the features of the experimental and theoretical investigation of the density of electron states (DOS), and we develop an approach to the analysis of DOS obtained theoretically and experimentally for noble transition metals, for example, the surfaces of polycrystalline samples of d-metals Au, Pd. In the future, it is planned to develop an approach to implement for the analysis of metallic nanoclusters and porous nanocluster films.

One of the most common and informative experimental methods for studying electronic properties is X-ray photoelectron spectroscopy (XPS) [1]. Its essence is in obtaining the distribution of the number of emitted photoelectrons depending on their binding energy, and this distribution describes the density of occupied electronic states in the material. It should be noted that photoelectrons can be collected from the depth no more than 2-3 nm or 10-20 monoatomic layers, so the method is sensitive only to the upper surface layer [1]. The analysis of publications concerning DOS, started from the earliest studies of the bulk state, and up to the modern calculations of nanoclusters, shows that the comparison of theoretical DOS with XPS spectra has for the most part a qualitative similarity. Generally, the comparison of volumetric DOS and XPS spectrum is not a comparison of two identical systems, since XPS is dealing with a surface, whose state is changed in comparison with the bulk structure of the material, even the idealy prepared surface – avoiding adsorption, oxidation, etc. It is necessary to compare with the experimental spectrum the theoretically calculated surface spectrum of the investigated material of that thickness and with those surface features specific for the sample and experiment.

A feature of nanoscale material modeling techniques is presence of both quantum (ab initio) and many-body approaches. For structures with a big (>10^3) number of atoms very powerful approach is density functional theory (DFT) with pseudopotential. In this work we use plane-wave basis set and pseudopotential DFT technique, with geometric optimization, implemented in PW DFT code Quantum Espresso [2]. Calculations was performed using resources of NRNU MEPhI high-performance computing center.

### 1. DFT DOS calculation for d nobble metals, particularly Au and Pd.

The properties of noble metals are due to their d-band filled and localized near the Fermi level. The width of the d-band, it position relative to the Fermi level, spin-orbit splitting is DOS characteristics which cause physical properties, and it all can be measured experimentally by the XPS spectrum [4].

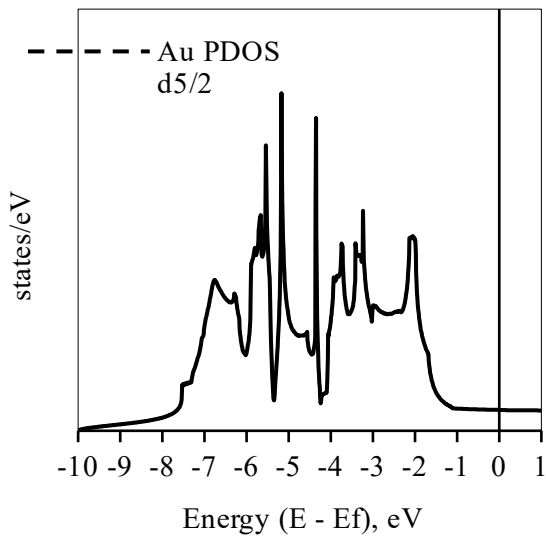
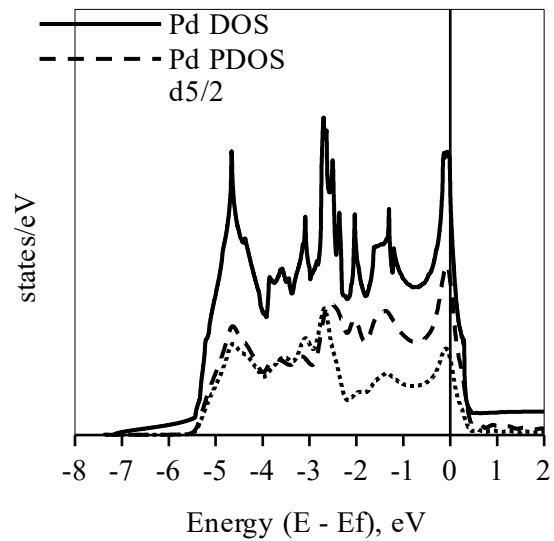

**Fig. 1a.** Total and d-DOS of bulk Au            **Fig. 1b.** Total and d-DOS of bulk Pd

The DFT study of the electronic structure of bulk gold and palladium was provided in DFT GGA (PBE) approximation [3] of the Ex, for the noncollinear spin orientation with the spin-orbit interaction and nonmagnetic state. The interaction of valence electrons with the core is described with the full-relativistic ultrasoft Vanderbild pseudopotentials [5]. The obtained DOS show a more accurate correspondence to the experimental data, spin-orbit splitting, rather than the results of DOS from databases for bulk metals [6,7]. Figures 1a, 1b show the calculated total and d-projected DOS. *PBE)*., This approximation of the DFT correctly describes the valence band, its distance to the Fermi level and the position of the peaks, the spin-orbit splitting of $d_{3/2}$, $d_{5/2}$ on atomistic scale, and was therefore also used for modeling nanoscaled surfaces.

## 2. Metal Au, Pd surfaces modeling

The structure with which photoelectrons are analyzed in XPS of these noble metals is a polycrystalline surface with a thickness of about 1 nm [1]. An arbitrarily cleaved surface consists of sections of surfaces of different crystalline faces. Like the growth of fcc clusters and spontaneous surface formation, it is most likely that the most closely packed faces with the lowest surface energy are [111], 100 ([200]). Analysis of diffraction patterns of polycrystalline gold shows that the greatest intensity is given by the faces [111]: [200]: [220] in the ratio ~ 2:1:0.8. Similar pictures are observed for thin films, surfaces, fcc-nanoclusters Au, Pt, Pd. [8].

In the XPS, the shape of the spectral line is related to the density of electronic states n (E) as:

$$I(E) = I_0 \int I_{DS}(E - E')n(E')G(E', \sigma)dE' \qquad (1)$$

G (E'; σ) is the total instrumental broadening, which is described by the Gaussian function, and the Dons-Schünich $I_{DS}$-function [9]. The fluxes of photoelectrons from each part of the surface (with information on the occupied electronic states in it) are summed in the detector, therefore, because of the transformation (1) is linear, the spectrum can be represented as a superposition of the spectra I ($n_i$ (E)) from the sites of the most probable hkl configuration - [111], [200], [220] weighed with Ci corresponding to the peaks of X-ray diffraction.

$$I(E) = \sum_{i=[hkl]} C_i \cdot I(n_i(E)) \qquad (2)$$

An infinite film with a thickness of 1 nm, cut in the [hkl] direction in a fcc crystal, is considered. To solve the periodic problem for such a structure, such a Bravais pseudo-lattice is constructed so that the periodicity in all directions is preserved, but the films do not interact. At the same time, the upper surface of the film remains free, and the lower surface is part of the bulk. In the absence of part of the bonds, the atoms on the surface are rearranged so as to minimize the energy of the system, on the other hand, the position of the atoms inside the film must be equal to the bulk one.

The Bravais pseudo-lattice primitive cell is composed of 2×2 minimal cells in the *XY* plane and 5 atomic layers in *Z* to provide a layer thickness of 1 nm and the motion of atoms relative to each other during geometric optimization, the interatomic distances in the first 2 layers are fixed, the films are separated by a 1.5 nm. Hexagonal and tetragonal Bravais lattices were constructed for [111] and [200] [220] films, respectively. Geometric optimization of these structures was performed using the BFGS algorithm (Broyden-Fletcher-Goldfarb-Shanno) [10] implemented in the Quantum Espresso code.

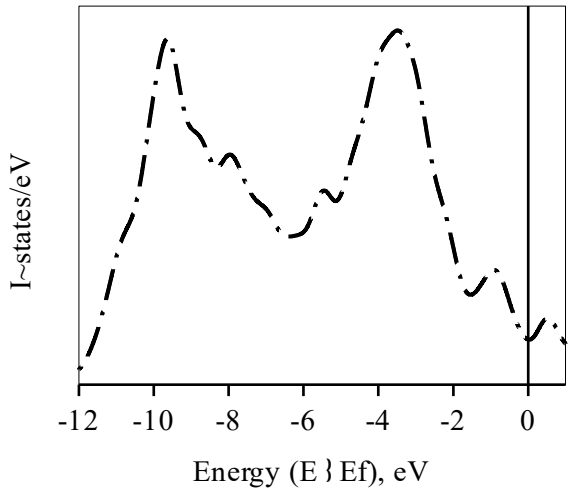

**Fig. 2a**. Smeared DOS of Au model surfaces.

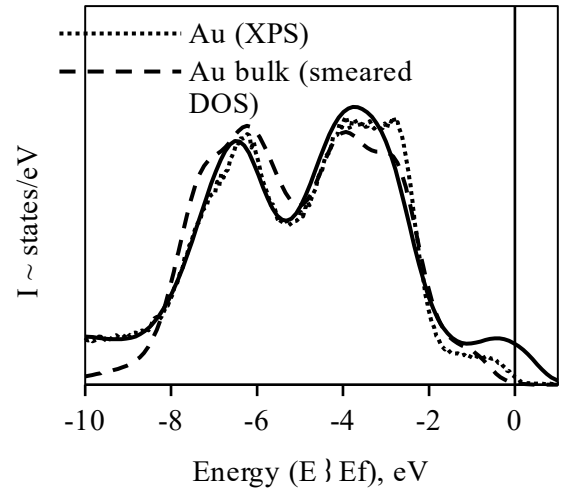

**Fig. 2b.** . Au valence band lines: experimental XPS, calculated from model surface DOS, and from bulk DOS..

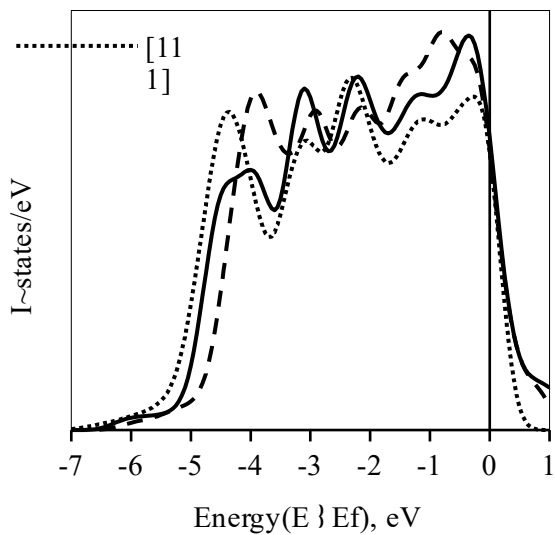

**Fig. 3a**. Smeared DOS of Pd model surfaces.

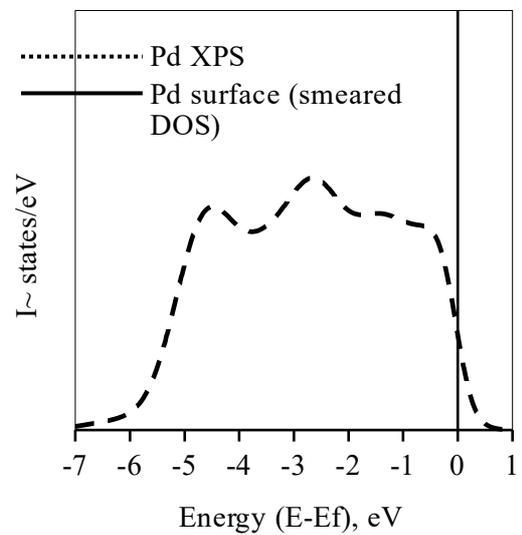

**Fig. 3b** . Pd valence band lines: experimental XPS, calculated from model surface DOS, and from bulk DOS.

In Fig. 2a showed the smeared DOS of model areas of the surface $I_{[hkl]i}$ and in Fig. 2b their superposition according to rule (2) with instrumental broadening, describing the XPS of the valence band of gold. The same for palladium is shown in Fig. 3a, 3b. It can be seen that the model of a surface more accurately describes the filled electronic states of the valence band (XPS) than the model of a bulk sample.

Thus, this approach reveals the suitability for modeling the surface DOS in the aggregate with an experimental study, and will be used for the analysis of nanoclusters and porous films formed from them.

**Acknowledgements**

This work was financial supported by the Russian Federation President Grant to support young scientists (№ 14.Y30.17.2948-MK)

**References**

1. Physical foundations of methods for studying nanostructures and solids. Ed. V.D. Bormann. NRNU MEPhI, 2014

2. Paolo Giannozzi; at all. "QUANTUM ESPRESSO: a modular and open-source software project for quantum simulations of materials". Journal of Physics: Condensed Matter. 21 (39): 395502 (2009)

3. John P. Perdew, Kieron Burke, and Matthias Ernzerhof. Phys. Rev. Lett. 77, 3865.

4. Electronic structure, correlation effects and physical properties of d- and f- metals and they compounds. V.Yu.Irkhin and Yu.P.Irkhin. Cambridge international science publishing, 2007

5. http://www.quantum-espresso.org/wp-content/uploads/upf_files/

6. NIMS Material database http://mits.nims.go.jp/index_en.html

7. NIST Standard reference database http://webbook.nist.gov/chemistry/

8. Mohsen Asadi Asadabad, Mohammad Jafari Eskandari (2015) Transmission Electron Microscopy as Best Technique for Characterization in Nanotechnology, Synthesis and Reactivity in Inorganic, Metal-Organic, and Nano-Metal Chemistry, 45:3, 323-326

9. S. Doniach and M. Sunjic, J. Phys. C 3, 284 (1970)

10. C.C.Broyden, J.Inst.Math.Appl., 6 (1970) 64 . R.Fletcher, Comput.J., 13(1970) 317. D.Goldfarb, Math.Comput., 24 (1970) 23. D.F. Shanno, Math.Comput., 24 (1970) 647